\documentclass[preprint,showpacs,preprintnumbers,amsmath,amssymb]{revtex4}
 
\usepackage{graphicx}
\usepackage{dcolumn}
\usepackage{bm}

\renewcommand{\vec}[1]{\boldsymbol{#1}}

\begin{document}

\preprint{APS/123-QED}

\title{Three Dimensional Evolution of a Relativistic Current Sheet : \\
Triggering of Magnetic Reconnection by the Guide Field}

\author{S. Zenitani and M. Hoshino}
\affiliation{
Department of Earth and Planetary Science, University of Tokyo,
7-3-1, Hongo, Bunkyo, Tokyo, 113-0033 Japan;
zenitani@eps.s.u-tokyo.ac.jp
}


\begin{abstract}
The linear and non-linear evolution of
a relativistic current sheet of pair ($e^{\pm}$) plasmas
is investigated by three-dimensional particle-in-cell simulations.
In a Harris configuration, it is obtained that the magnetic energy
is fast dissipated by the relativistic drift kink instability (RDKI).
However, when a current-aligned magnetic field
(the so-called ``guide field'') is introduced,
the RDKI is stabilized by the magnetic tension force
and it separates into two obliquely-propagating modes,
which we call the relativistic drift-kink-tearing instability (RDKTI).
These two waves deform the current sheet so that
they trigger relativistic magnetic reconnection at a crossover thinning point.
Since relativistic reconnection produces a lot of non-thermal particles,
the guide field is of critical importance to study the energetics of a relativistic current sheet.
\end{abstract}

\maketitle


It is widely believed that
plasma heating and particle acceleration occur
in a wide variety of plasma regions that contain magnetic fields.
A current sheet structure where the magnetic field polarity changes its direction
is one of the most fundamental structures among them.
Importantly, when two groups of magnetic field lines with opposite polarities meet each other
around the current sheet, magnetic reconnection takes place
and then it explosively dissipates the magnetic energy into the kinetic energy of plasmas.
In fact, reconnection is accepted as a main player in stellar and solar flares \citep{parker} and
storms in the Earth's magnetosphere \citep{dungey}.
%
The theory of reconnection has often been studied in
a current sheet with anti-parallel field lines but
it is possible that field lines are somehow ``twisted''
and so the system with a finite amplitude of a current-aligned magnetic field
(the so-called ``guide field'') has recently been investigated
by three-dimensional (3D) simulations \citep{drake03,ricci03,silin03,prit04,tanaka04}.
Magnetic reconnection processes are also important
in high-energy astrophysical contexts;
the jets from active galactic nuclei \citep{dimatteo,les98,lar03},
pulsar wind  \citep{coro90,kirk03}
and probably gamma-ray bursts \citep{dr02,drs02}.
Particularly in the Crab Nebula,
it has been a long-standing problem (the so-called ``$\sigma$-problem'')
how originally Poynting-dominated plasmas \citep{arons74}
are converted into kinetic-dominated plasmas \citep{kc84}
in the relativistic outflow from the neutron star.
The most promising solution is relativistic magnetic reconnection \citep{blackman}
in the striped current sheets \citep{coro90}.

However, reconnection processes in such
a relativistic hot $e^{\pm}$ plasmas
(the plasma temperature $T$ is larger than the rest mass energy: $T \gg mc^2$)
has been poorly understood.
Recently, 2D particle simulations of relativistic magnetic reconnection
in such a relativistic current sheet (hereafter RCS)
were carried out \citep{zeni01,claus04}.
One important point is that a lot of magnetic energy can be converted
into non-thermal energy of plasmas due to the enhanced dc acceleration 
around the X-type region.
By the way, a RCS is extremely unstable to
the relativistic drift kink instability (RDKI) \citep{zeni05},
a relativistic extension of the drift kink instability (DKI) \citep{zhu96,prit96,dau98},
whose wave vector is normal to 2D reconnection plane.
Contrary to the relativistic reconnection,
the RDKI mainly dissipates magnetic energy into plasma heat
rather than non-thermal energy.
In addition, the effect of the guide field in a RCS has never been investigated in 3D/2D.

We have performed 3D particle-in-cell  simulations
to explore linear/nonlinear evolution of a RCS,
firstly taking into account the guide field effect.
The system consists of $256^3$ grids.
Periodic boundaries in the $x$, $y$ and $z$ directions and
with double current sheets are considered in the $z$ direction.
The half width of the current sheet $\lambda$ is set to 10 grids
so that the boundaries are located at $x = \pm 12.8 $,
$y = \pm 12.8$ and $z = \pm 6.4$ in unit of $\lambda$.
Also, time is normalized by the light transit time $\tau_c = \lambda/c$.
We take a relativistic Harris model as the initial current sheet configuration \citep{kirk03}.
The magnetic field and
their distribution functions are described by
$B_x = B_0 \tanh(z)$, $B_y =\alpha B_0$,
$f_{\pm} \propto n(z) \exp[-\Gamma_\beta\{\varepsilon - \beta_\pm mcu_y\} / T ] +
n_{bg}\exp (-\varepsilon / 0.1T )$ and
$n(z) = n_0 \cosh^{-2} (z)$,
where $B_0$ is the magnitude of anti-parallel component of the magnetic field
in the lobe (the background region),
$\alpha$ is the relative amplitude of the guide field to the reversed field,
$n_0$ is the number density of plasmas in the current sheet,
$n_{bg}$ is the number density of plasmas in the lobe,
$\beta_\pm = v_{\pm}/c$ are the drift velocities for each species;
$\beta_{+} = + \beta$ for positrons,  $\beta_{-} = - \beta$ for electrons,
$\Gamma_\beta = [1-\beta^2]^{-1/2}$,
$\varepsilon$ is the particle energy and
$\bm{u}$ is the four velocity of $\bm{u}= [1-(v/c)^2]^{-1/2} \cdot \bm{v}$.
We investigate two cases of $\alpha=0$ (Run A) and $\alpha=-0.5$ (Run B).
We set $T = mc^2$ and $\beta = 0.3$.
The total number of super particles is $5\times 10^8$
and we set the plasma density to $n_{0}\sim 80$ (pairs) and $n_{bg} \sim 4$ (pairs) per one grid
so that $n_{bg}/n_{0} = 0.05$.
The total energy is conserved within an error of $0.5\%$
throughout the simulation runs.

\begin{figure*}
\begin{center}
\includegraphics[width={1\columnwidth},clip]{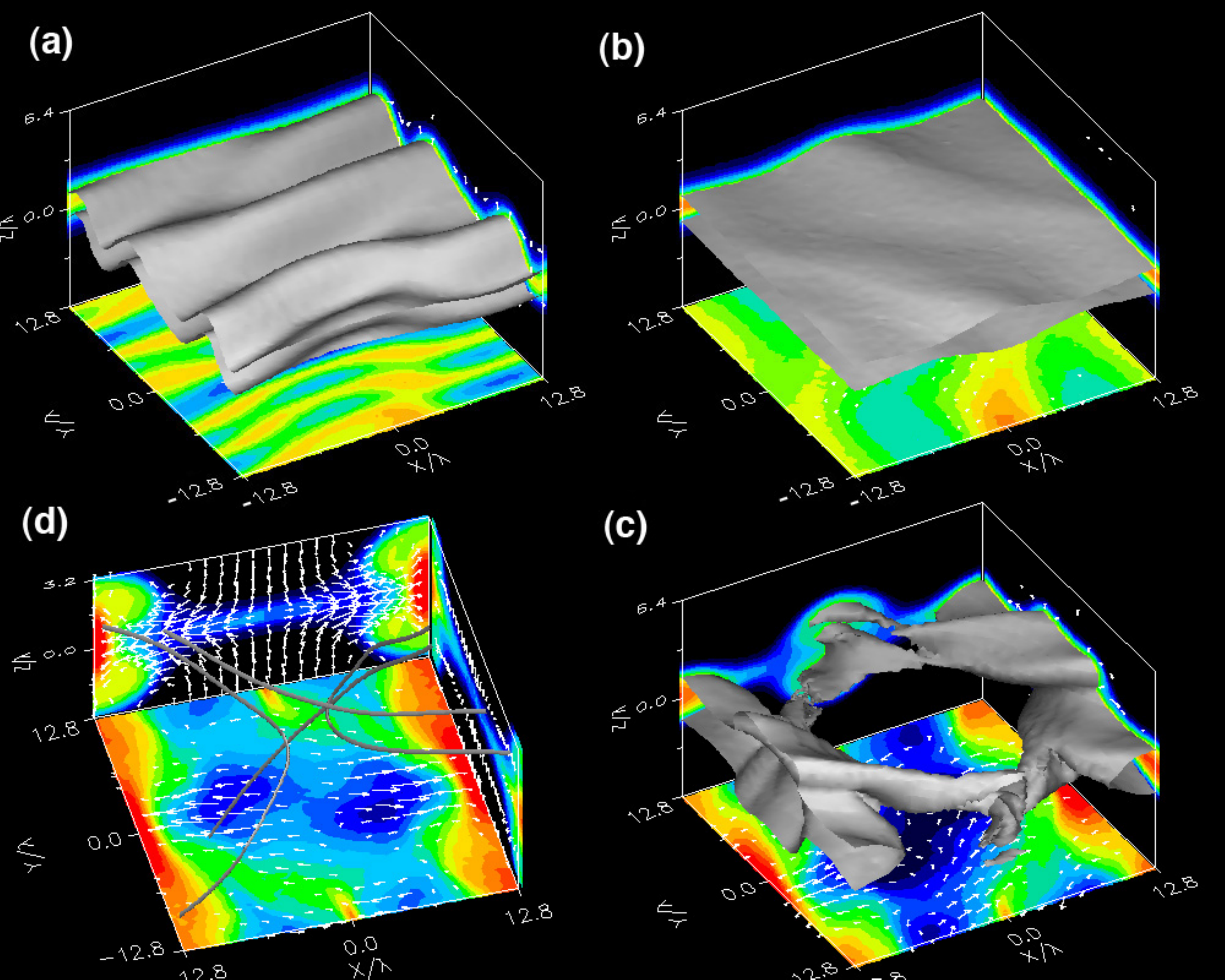}
\caption{\label{fig:4plot}(color)
(a) Snapshot of the current sheet in Run A ($ \alpha =0$) at $t=80$.
Two gray surfaces show $n=2/3n_0$.
The plasma density at the neutral plane ($z=0$) is projected into the bottom roof,
with color from black (empty)  to red (dense; $n\sim1.2 n_0$).
The white arrow shows plasma flow.
The light speed ($v=c$) is projected to the length of $4$.
(b) Snapshot of the current sheet in Run B
with a guide field configuration ($\alpha = - 0.5$) at $t=120$.
Two gray surfaces show $n=2/3 n_0$ and
the plasma density under the neutral plane ($z={-1}$) is presented
in color in the bottom roof.
(c) Snapshot at $t=170$ in Run B.
(d) Another snapshot at $t=170$ in Run B, with typical magnetic field lines.
Three projections show cross-sections at $x,y,z=0$.
}
\end{center}
\end{figure*}

\begin{figure}
\begin{center}
\includegraphics[width={0.7\columnwidth},clip]{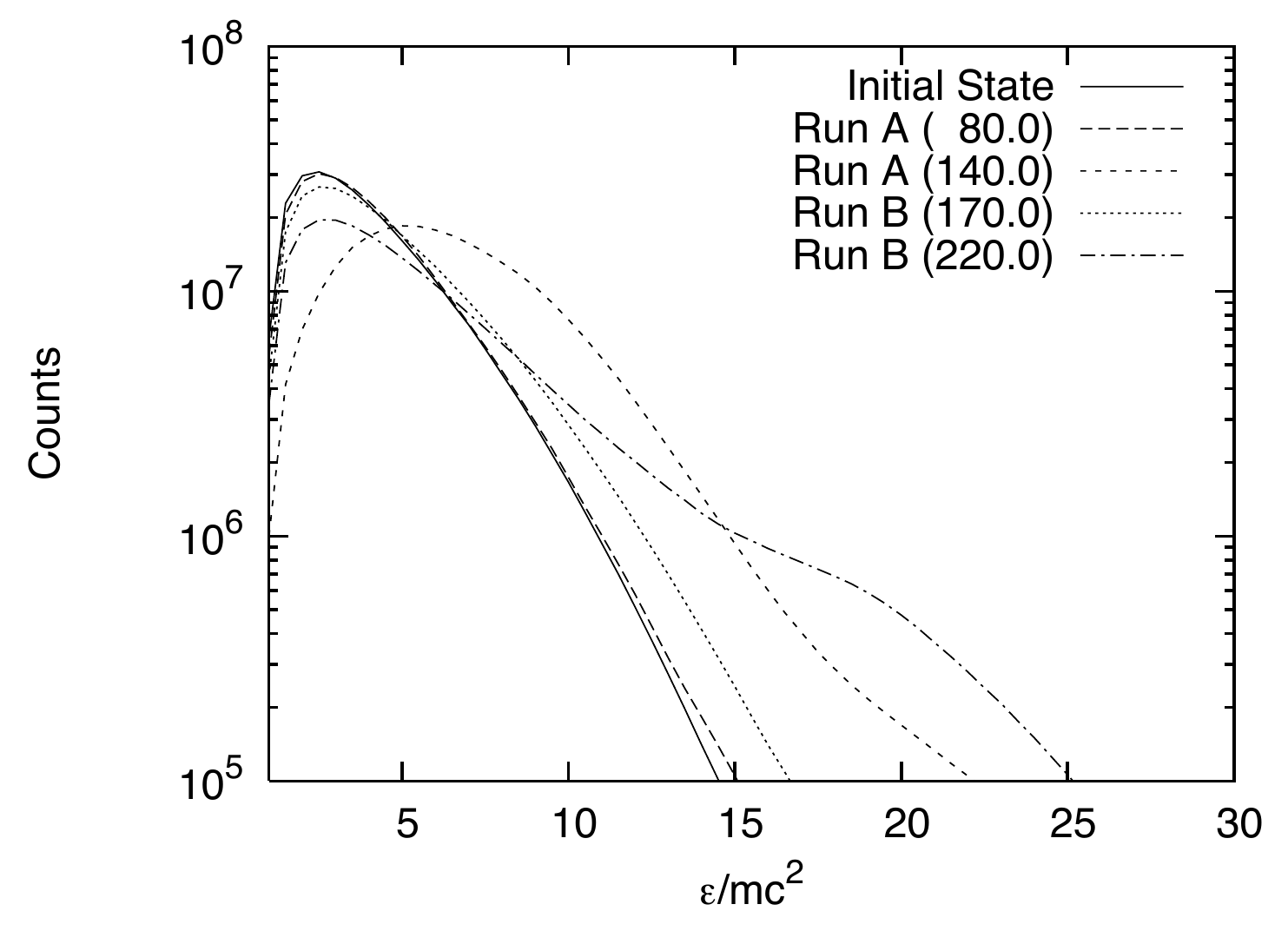}
\caption{\label{fig:espec}
Energy spectra observed in Runs A and B.
The initial state (the same in both two runs) and two typical stages for each runs are selected.
Time is normalized by $\tau_c$ and particle energy is normalized by $mc^2$.
}
\end{center}
\end{figure}

First, we present a result of Run A without guide field.
A snapshot at $t=80$ is presented in panel (a) in Fig.~\ref{fig:4plot}.
The system evolution is similar to that of the 2D RDKI \citep{zeni05},
including profiles of perturbed physical properties.
The wave-number of the most dominant mode is $k_y \sim 0.74$ (mode 3)
and its growth rare is $\omega_i/\Omega_c=0.02$,
where $\Omega_c$ is the gyro-frequency;
$\Omega_c=\omega_c/\gamma=(eB_0)/(\gamma mc)$,
$\gamma$ is the Lorenz factor for typical particle energy.
In the 2D work, the growth rate has
its peak around $k_y \sim 0.7 - 0.8$ at $\omega_i/\Omega_c=0.03-0.04$.
After this stage, the RCS becomes folded, turns into the non-linear stage and then
mixed into the turbulent state.
Because of the limited simulation height in $z$,
some plasmas are also mixed across the $z$ boundaries.
Finally, more than 83\% of the magnetic energy is dissipated into the particle energy.
%
%
Fig.~\ref{fig:espec} shows energy spectra of 3D runs.
For Run A, three stages ($t=0, 80, 140$) are presented.
In later time of $t=140$,
one can recognize a high-energy tail
due to the particle acceleration inside folded current sheet.
More importantly, plasmas are heated
throughout the late-time mixing stage of the RDKI.
The transportation across the $z$ boundaries
provides no significant change between 2D and 3D in the spectra.
We think that the observed growth rate becomes slower than that of 2D
due to the limited system size in $z$.
Note that it is still faster than that of 2D reconnection ($\omega_i/\Omega_c=0.011$).
Since the RDKI grows faster when $T \gtrsim mc^2$ \citep{zeni05},
we conclude that the RDKI is a dominant process in a RCS.
Again, magnetic dissipation and plasma heating by the RDKI
would be the main signature of a RCS.

Next, we show results of Run B
with a finite amplitude of the guide field $B_y = - 0.5B_0$.
Snapshots at two stages ($t=120$ and $170$) are presented
in panels (b), (c) and (d) in Fig.~\ref{fig:4plot}.
Instead of the RDKI,
a flute-like mode whose wavevector is $\vec{k}_1 = (0.25,0.25)$,
is observed on the upper side of the RCS at $t =120$ in Fig.~\ref{fig:4plot} (b).
Similar flute mode is observed on the lower side, too,
but its wavevector is in different direction: $\vec{k}_2= (0.25,-0.25)$.
In order to take a clear look at this lower-side mode,
we present cross-sections at $z=-1$ in panels (b) and (c).
Snapshots at $t=170$ are presented in panels (c) and (d) in Fig.~\ref{fig:4plot}.
Due to the two flute modes, the RCS is so modulated that
a plasma density hole appears at the center.
In panel (d), cross-sections of the RCS at $x,y,z=0$
and typical magnetic field lines are presented.
Magnetic reconnection occurs around the center of the simulation box.
The speed of reconnection jets is up to $0.6c$ and
the outflow plasmas are transported into the O-point(s) around $(x, y) \sim (\pm 12.8, 0)$.
Then, since magnetic field mainly consists of the guide field component $B_y$ inside the RCS,
they are dissipated along the O-line region around $x=\pm 12.8$.
The dense region around $(x, y) \sim (0, \pm 12.8)$ is
the remnant of the thick point of the RCS,
where two obliquely-propagating modes are linked.
%
Energy spectra in Run B are also presented in Fig.~\ref{fig:espec}.
The oblique mode itself has few effect on the spectra, however,
after relativistic reconnection starts,
strong enhancement of the non-thermal tail is observed
due to the particle acceleration around the X-type region \citep{zeni01,claus04}.

Along with 3D work,
we have studied an effect of the guide field
using several sets of 2D particle simulations ($|\alpha|=0, 0.25, 0.5, 1.0$).
It is obtained that reconnection in the $xz$ plane grows slower when larger $|\alpha|$ is set,
while we were unable to observe the the RDKI in the $yz$ plane
when $|\alpha|$ exceeds some critical value ($|\alpha| \gtrsim 0.5$).
We found that this is due to the magnetic tension effect.
When $\alpha=0$, magnetic field lines are always parallel to the wavefronts of the instability
and so they cannot be bent (no tension force)
but when the guide field is introduced,
the RDKI will be opposed by the tension force of current-aligned field lines.

\begin{figure}
\begin{center}
\includegraphics[width={\columnwidth},clip]{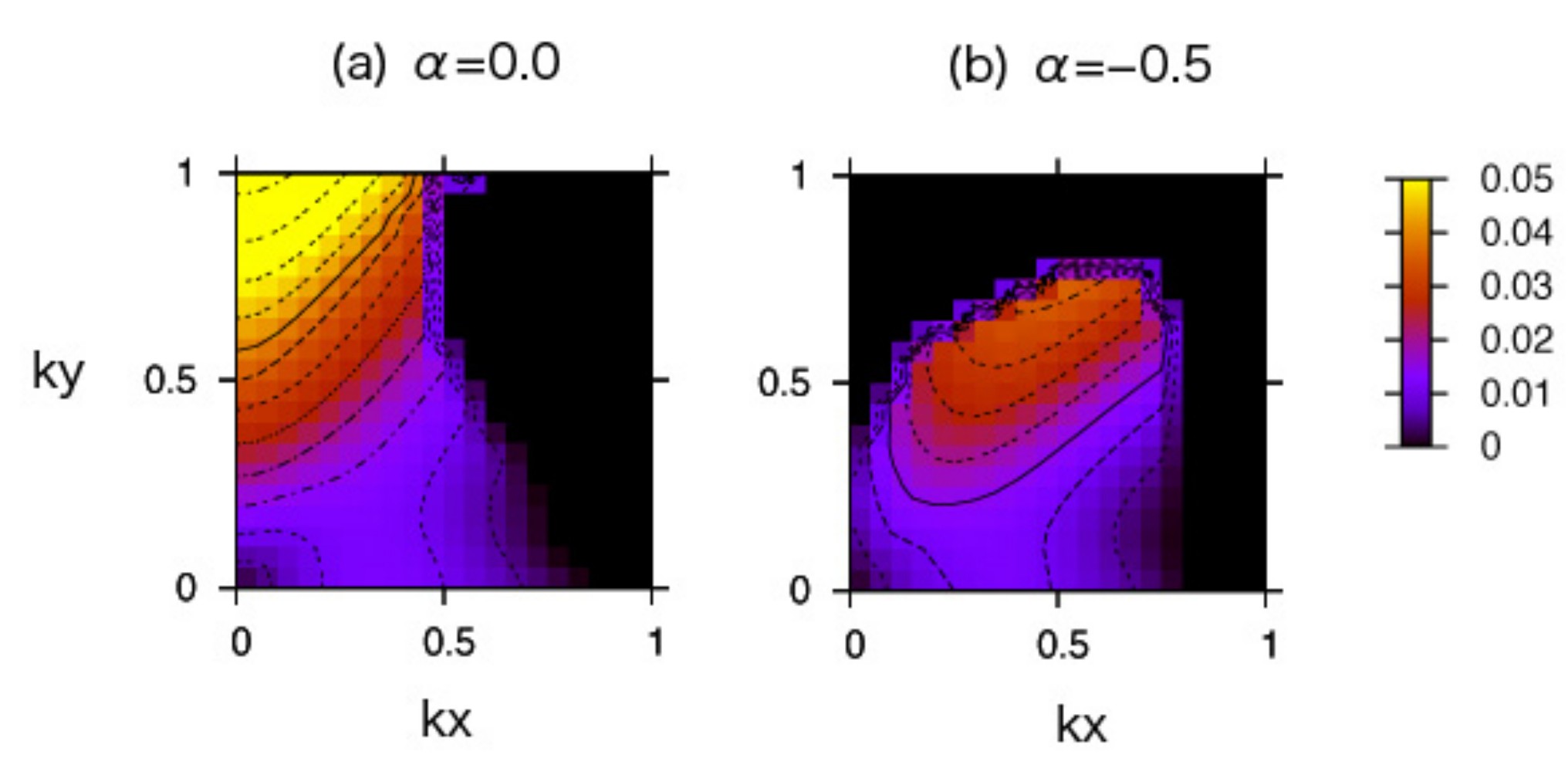}
\caption{\label{fig:map}
(color online)
Growth rate ($\omega_i/\Omega_c$) of the instabilities
in wave-vector spaces of $(0\le k_x \le 1, 0 \le k_y \le 1)$.
(a) $\alpha=0$ and
(b) $\alpha=-0.5$.
}
\end{center}
\end{figure}

\begin{figure}
\begin{center}
\includegraphics[width={\columnwidth},clip]{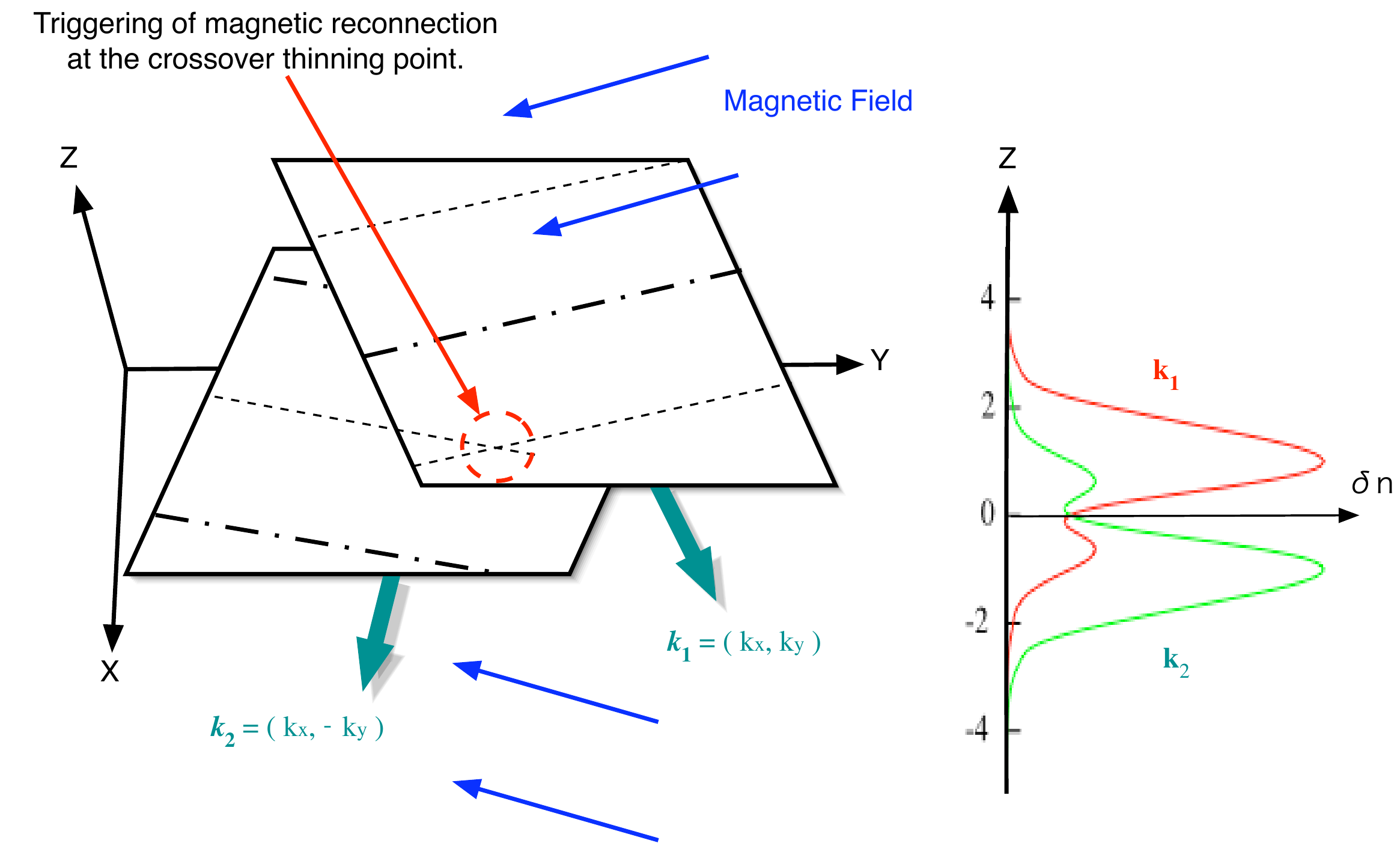}
\caption{\label{fig:scheme}
Left: A schematic illustration of obliquely-propagating modes and
the triggering mechanism of magnetic reconnection.
Right: eigen profiles of the relativistic drift-kink-tearing modes.
A gray line stands for $\vec{k}_1=(0.25, 0.25)$ and
a light gray line shows its partner: $\vec{k}_2=(0.25,-0.25)$.
}
\end{center}
\end{figure}

In order to study the obliquely-propagating modes in Run B,
we investigate eigen functions and their linear growth rate of instabilities in the RCS.
The relativistic two-fluid equations are linearized,
assuming that perturbations are given by $\delta f \propto f(z) \exp(ik_x x + ik_y y - i \omega t)$.
The obtained growth rate ($\omega_i/\Omega_c$)
are presented in contour maps in Fig.~\ref{fig:map}
for two cases of $\alpha=0, -0.5$.
In these maps, reconnection (tearing mode) is plotted along the $k_x$ axis
and the RDKI is plotted along the $k_y$ axis.
We mention that
all of the obtained modes are purely growing
and that growth rate for $(k_x, k_y)$ are equal to the rate for $(|k_x|, |k_y|)$
because of the symmetry.
In Fig.~\ref{fig:map} (a),
the RDKI or its neighbor of the relativistic drift sausage instability (RDSI)
has the maximum growth rate in $k_x=0$ while
we obtain the dominant mode at $(k_x, k_y) \sim (0, 0.74)$ in the 3D simulation.
We note that these instabilities
are dumped by the kinetic effect
in shorter wavelength of $|k| \gtrsim 1$,
where the fluid theory is not valid.
Because of the similarity of the eigen profiles and growth rates,
it is natural to say that the oblique modes are intermediate mode
between the RDKI/RDSI and the tearing instability.
Hereafter we shall call them the ``relativistic drift-kink-tearing instability'' (RDKTI) \citep{zeni08},
the oblique extension of the RDKI/RDSI
driven by $\vec{k}$-aligned component of the current.
Since the RDKTI mode is weakly stabilized by the field lines and
since its driving force is weaker,
the RDKTI grows slower than relevant RDKI/RDSI in a RCS with exact anti-parallel fields.
When we set larger $|\alpha|$, the RDKI/RDSI along the $k_y$ axis
become slower due to the magnetic tension of stronger guide field.
Instead, two branches of the RDKTI for $\vec{k}=(k_x,\pm k_y)$ become dominant,
or, in other word, the RDKI/RDSI separates into two branches of oblique RDKTI waves.
Fig.~\ref{fig:map} (b) shows growth rate in case of $\alpha=-0.5$.
As obtained by supplemental 2D runs,
the RDKI along the $k_y$ axis is stabilized by the magnetic tension.
The right panel in Fig.~\ref{fig:scheme} shows eigen profiles of
density perturbation for $\vec{k}=(0.25, \pm 0.25)$.
They are highly asymmetric in $z$;
a mode for $\vec{k}_1=(0.25, 0.25)$ has its peak in the upper side of the RCS
while the other one for $\vec{k}_2=(0.25, -0.25)$ has its peak in the lower side.
Such asymmetry is explained by ``twisted'' effect of background magnetic fields.
In case of the mode for $\vec{k}_1$,
the magnetic field is quasi-perpendicular to $\vec{k}_1$
in the upper side of the RCS so that its tension can be neglected.
On the contrary, the field is quasi-parallel to $\vec{k}_1$ in the lower side,
so that the mode is opposed by strong magnetic tension.
The RDKTI for $\vec{k}_2$ is
an upside-down mirror of the RDKTI for $\vec{k}_1$ because of the physical symmetry.
As a result, we observe two dominant RDKTI modes:
one for $\vec{k}_1$ in the upper side and the other for $\vec{k}_2$ in the lower side of the RCS.
As larger $|\alpha|$ is set,
an angle between two oblique wave-vectors (such as $\vec{k}_1$ and $\vec{k}_2$) becomes wider
in accordance with the lobe magnetic field lines.
In such cases, the RDKTI grows slower because of stronger magnetic tension.

The secondary relativistic reconnection is triggered by the RDKTI in its non-linear stage.
Since two major RDKTI modes are in different direction,
the deformed RCS become very thin at the cross-over point
where both the upper and the lower RDKTI wave press the RCS.
This mechanism is illustrated in Fig.~\ref{fig:scheme}.
In addition to this enhanced thinning effect,
accompanying electric fields $E_y$ and the heated plasma due to the RDKTI
leads to the triggering of magnetic reconnection.
Once reconnection broke up, it continue to grow and
then particle acceleration takes place near the reconnecting region.
It produces more nonthermal particles
due to the direct particle acceleration in the reconnection region
\citep{zeni01,claus04}.
Actually there are two crossover thinning points --
one is around $(x,y)\sim(0,0)$ and
the other is around $(x,y)\sim(\pm12.8,\pm12.8)$.
However, reconnection jets from the former point
are finally transported into the latter point
so that the reconnection does not dominate at the latter point.
The speed of the reconnection jets is finally up to $0.78c$ at later stage.

The electromagnetic energy ($E_{EM}$)
is dissipated by the rate of $\frac{d}{dt}E_{EM} \sim - 4 v_t L_x L_y B_0^2/8\pi$,
where $L_{x,y}$ are the system size and $v_t$ is
a typical dissipation speed \citep{dr02,kirk03}.
We obtain $v_t \sim 0.25c$ in Run A
and this value is consistent with
the $z$-displacement speed of the RDKI: $v_z=0.2-0.3c$ \citep{zeni05}.
In Run B, $v_t \sim 0.05c$ or larger.
The ratio of the total field energy ($E_{EM}$) to
the total particle energy ($E_{P}=\sum \gamma mc^2$)
is originally $1.0$ but it falls below $0.1$ at $t=160$ in Run A,
while the ratio evolves from $1.25$ to $0.7$ at $t=220$ in Run B.
Although this ratio extremely depends on the system-size,
it seems that the RDKI is favorable to produce kinetic-dominated plasmas.

Finally, we discuss our system size limitations.
The most dominant RDKTI has mode $(1,\pm 1)$ in Run B.
Since smaller-wavelength modes will be suppressed by the kinetic effect
and since longer-wavelength RDKTI modes has smaller growth rate in Fig.~\ref{fig:map} (b),
the obtained modes would be roughly correct and
as long as a pair of the RDKTI modes dominate,
the trigger mechanism of reconnection should be the same.
The global evolution of the RCS still remains unclear.
One possibility is that multiple reconnection regions evolve into one big reconnection region,
and then higher energetic particles will be produced in the wider acceleration region.
Another possibility is that many reconnection regions grow and that
the system becomes turbulent, which may contribute to statistical acceleration or thermalization.
It should be further investigated by
larger particle simulations or two-fluid relativistic MHD simulations.

Let us summarize this paper.
Considering the effect of the guide field,
we present the following two scenarios in a RCS.
One is plasma heating by the RDKI in an anti-parallel case without the guide field.
The other is non-thermal acceleration by secondary reconnection in a twisted case with the guide field.
In this case, magnetic reconnection is triggered by the coupling of two RDKTI waves,
which is separated from the RDKI/RDSI.
It should be noted that slight difference in magnetic field topology have great influence on
the destination of the magnetic energy: plasma thermal energy or non-thermal energy.
%

The authors are grateful to Dr. T. Yokoyama and Dr. A. Miura for fruitful discussions.
This work was supported by the facilitates of
JAXA and the Solar-Terrestrial Environment Laboratory, Nagoya University.

%

\clearpage

\clearpage

\clearpage

\clearpage

\end{document}